# A magnetic monopole nanoantenna

Benoît Reynier[1], Xingyu Yang[1], Bruno Gallas[1], Sébastien Bidault[2] and Mathieu Mivelle[1*]

[1] Sorbonne Université, CNRS, Institut des NanoSciences de Paris, INSP, 75005 Paris, France

[2] Institut Langevin, ESPCI Paris, Université PSL, CNRS, 75005 Paris, France

*Corresponding author: mathieu.mivelle@sorbonne-universite.fr

ORCID:0000-0002-0648-7134

Magnetic monopoles are hypothetical particles that, like electric monopoles which generate electric fields, are at the origin of magnetic fields. Despite many efforts, to date, these theoretical particles have yet to be observed. Nevertheless, many systems or physical phenomena can be related to magnetic monopole behavior. Here, we propose a new type of photonic nanoantenna behaving as a radiating magnetic monopole. We demonstrate that a half-nanoslit in a semi-infinite gold layer generates a single pole of an enhanced magnetic field at the nanoscale and that this single pole radiates efficiently in the far field. This original antenna concept opens the way to a new model system to study magnetic monopoles, to a new source of optical magnetic field to study the "magnetic light" and matter coupling, and allows potential applications at other frequencies such as magnetic resonance imaging.

Magnetic monopoles are hypothetical particles that carry a single magnetic pole, as opposed to all other magnetic objects that come in pairs of magnetic poles. The concept of magnetic monopoles was first proposed by Paul Dirac in the 1930s,[1] and has since been the subject of much theoretical and experimental research in the following decades.[2]

In many ways, magnetic monopoles are similar to electric charges. Just as electric charges are the fundamental building blocks of electric fields, magnetic monopoles are supposed to be the fundamental building blocks of magnetic fields (figure 1). However, unlike electric charges, that exist as isolated entities, magnetic poles always come in pairs (north and south). Indeed, despite many experimental investigations, magnetic monopoles have yet to be definitively observed in Nature. Many physicists believe that they must exist in order to explain certain phenomena, such as the observed quantization of electrical charges.[2] In recent years, there have been several proposed methods[3,4] to investigate magnetic monopoles, including the use of high-energy particle accelerators, cosmic ray detectors, and the quest for magnetic monopole "tracks" in certain materials.

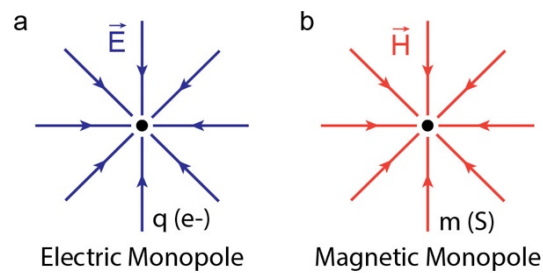

*Figure 1. Description of electric and magnetic monopoles. a) A negative electric point charge generates electric field lines oriented in the direction of the charge. b) A south magnetic point pole creates magnetic field lines in the direction of the pole.*

Although magnetic monopole particles are yet to be observed experimentally, for many years, physicists have been trying to develop systems behaving like magnetic monopoles or have used the concept to explain the behavior of certain materials. This is the case, for instance, in topological insulators,[5] spin ice materials,[6] high-temperature superconductors,[7] during heavy ion collisions,[8] in quantum Hall systems,[9] skyrmions,[10] liquid crystals,[11] Weyl semimetals,[12] or chiral magnets,[13] to name a few.

To contribute to this endeavour, this paper proposes a new type of plasmonic nanoantenna, based on a half-nanoslit in a semi-infinite gold layer and behaving as a magnetic monopole. We demonstrate that our plasmonic nanostructure develops a single and enhanced magnetic hot spot at the nanoscale, which is locally free of any electric field. Furthermore, this enhanced energy density of optical magnetic field is made of only one orientation of the magnetic field, constituting a single pole, as opposed to plasmonic magnetic dipoles carrying two opposite poles concentrating the magnetic field. Finally, we describe how this monopole oscillates in time at the excitation frequency, alternating north-south orientations, allowing this antenna to efficiently radiate electromagnetically in the far field from this single pole, similarly to how an electric monopole antenna would behave.

The results presented here are of great significance for several reasons. Firstly, this nanoantenna can be used as a model system to study the behavior of magnetic monopoles. Also, by its ability to concentrate, isolate and enhance the optical magnetic field at the nanoscale, this type of nanostructures opens new possibilities for manipulating the coupling between magnetic light and matter.[14-23] In particular, research topics where the optical magnetic field plays a major role, such as the interactions between light and chiral matter,[24-26] electrically forbidden photochemical processes,[27] or nonlinear photon avalanche processes involving the magnetic transitions of lanthanide ions,[28] will benefit directly from this type of system. Finally, owing to its efficient far-field radiation and ability to concentrate the magnetic field locally in the near-field, and similarly to other types of photonic antennas, the extension of this new concept to GHz and MHz frequencies will profoundly impact certain technologies such as magnetic resonance imaging.

## Results and discussions

In electromagnetism, antennas can feature a broad range of features[29] in terms of directivity, bandwidth, or gain. In the visible range, these structures, called nanoantennas,[30] have been developed in different types of materials, dielectric[31-34] or metallic,[35,36] to cover a wide range of properties. Examples include fluorescence enhancement,[37] heat generation,[38] biosensing,[39] or directionality.[40] Many sophisticated antenna geometries have been used for these applications.[40,41] However, we know that some of the simplest nanoantenna geometries induce multipolar behaviors when they are excited by an optical field.[42] Figures 2a-c list some of these antennas. For instance, a gold nanorod (figure 2a) behaves like an electric dipole.[43] In particular, this behavior leads to a strong increase of the optical electric field (**E**) at both ends of the nanorod when it is excited at resonance by a light polarized with the electric field collinear to the long axis of the antenna (figure 2a). Also, by reciprocity and via Babinet's principle, we know that a nanoslit in a gold layer (figure 2b) behaves like a magnetic dipole.[44,45] In this case, the optical magnetic field (**H**) is enhanced at the ends of the nanoslit, at resonance, for an excitation polarization with a magnetic field that is collinear to the antenna's long axis (figure 2b). Similarly, it has been described that a metal half-nanorod attached to a semi-infinite layer of the same metal (figure 2c) behaves like an electric monopole,[46] this time with a single electric hot spot at the end of the half-nanorod (figure 2c). From then on, by reciprocity and via the same Babinet principle, one can wonder whether the electromagnetic behavior of a half-nanoslit in a semi-infinite layer of metal is similar to a magnetic monopole which would generate only one magnetic hot spot (one pole) at the end of it (figure 2d). This paper numerically demonstrates this property in comparison to the behavior of the magnetic dipole.

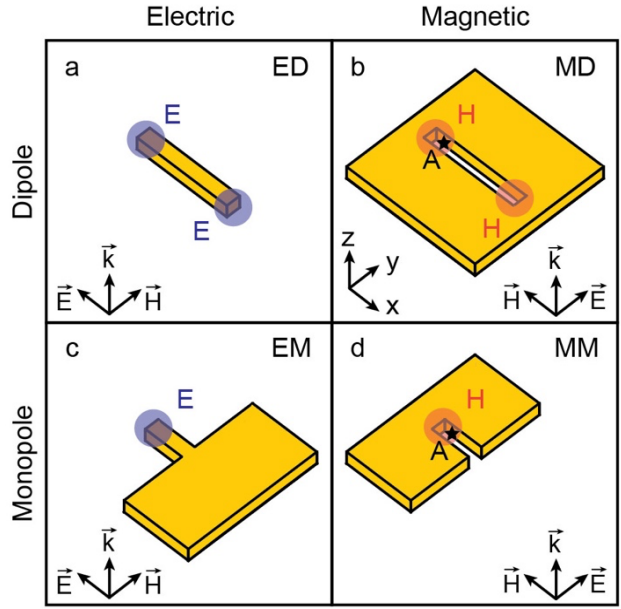

*Figure 2.* Schematic illustration of different plasmonic multipolar antennas. Design of dipolar a) electric, b) magnetic and monopolar c) electric, d) magnetic nanoantennas.

For this purpose, in this work, two structures are considered. One consists of a nanoslit opened in a thin layer of gold, hereafter called the magnetic dipole, shown in figure 2b, and the other is a half-nanoslit in a semi-infinite layer of gold, hereafter called the magnetic monopole and shown in figure 2d. These two nanostructures are both made in a 40 nm thick gold layer, and their groove width is 20 nm. The lengths of the latter have been chosen so that these photonic antennas resonate at a wavelength of 800 nm. These plasmonic structures are excited by a plane wave incident from the bottom side and polarized with the magnetic field collinear to their long axis (figure 2 b,d). Figure 3 represents the spectral response of these two magnetic antennas. In particular, it shows the increase of the local magnetic field at the surface of these antennas 10 nm from their ends and in the middle of the gold layer, as represented by the black star in the inset. From these spectral responses, it can be seen that these structures resonate at 800 nm wavelength and that the local magnetic field is increased by 12 and 6, respectively, for a dipole of length 140 nm and monopole of length 57 nm. In addition, we note a lower quality factor in the case of the magnetic monopole, indicating a faster dissipation of the energy of this antenna compared to the magnetic dipole, i.e. a high radiation efficiency.

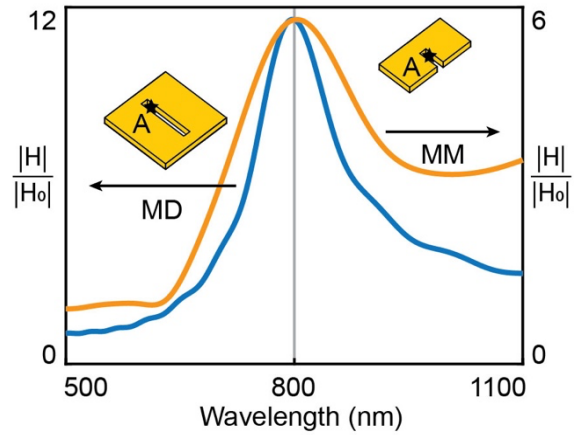

*Figure 3. Magnetic spectral responses. Near-field spectral responses of the increase in optical magnetic field for the magnetic dipole nanoantenna (blue curve) and for the magnetic monopole nanoantenna (orange curve) at the position symbolized by the black star, at a position 10 nm from the end of the nanoslits and in the Z-center of the gold nanostructures.*

To further detail the electromagnetic behavior of these antennas, the distributions of the electric and magnetic field increases in different planes of these structures are shown in Figure 4. In particular, the distributions of the electric (Fig. 4a,c,f,h) and magnetic (Fig. 4b,d,e,i) fields for the magnetic dipole (Fig. 4a-d) and the magnetic monopole (Fig. 4f-i) in the XY plane at the Z-center of the antennas (Fig. 4a,b,f,g) and the XZ plane at the Y-center of the antennas (Fig. 4c,d,h,i), are shown. As one can see, the magnetic dipole localizes the electric field inside the nanoslit and the magnetic field at both ends, this behavior is characteristic of a magnetic dipole antenna, as previously reported.[43] Furthermore, we can notice a substantial increase of the electric and magnetic fields amplitudes in the near field, particularly by a factor of 30 and 14, respectively. On the other hand, the spatial distribution of the optical fields around the magnetic monopole antenna is very different; we observe that the latter appears truncated compared to the magnetic dipole, with only one magnetic and one electric hot spot, at each end of the nanoantenna. Also, we can observe that the amplitudes of the fields in the case of the monopole are divided by two compared to the magnetic dipole. Everything happens as if we had only half of the magnetic dipole and, therefore, only one pole forming, in fact, a magnetic monopole.

To look more closely at this phenomenon, figures 4e and j describe the magnetic field lines in each of the antennas in an XZ plane at the Y-center of the nanostructures and at an arbitrary time t. As one can see, in the case of the magnetic dipole, where the magnitude of the magnetic field is the greatest, at the ends of the antenna, the orientation of the field lines is opposite (Figure 4e), with on one side an orientation towards positive Z (left) and on the other side towards negative Z (right). These two orientations can be seen as the north and south poles of the magnetic dipole antenna. On the other hand, in the monopole case, we have only one orientation of the magnetic field lines (figure 4j) at the position of the hot spot shown in figure 4i. Therefore, the antenna described in figures 4f-j behaves like an antenna with only one pole, i.e., like a magnetic monopole antenna.

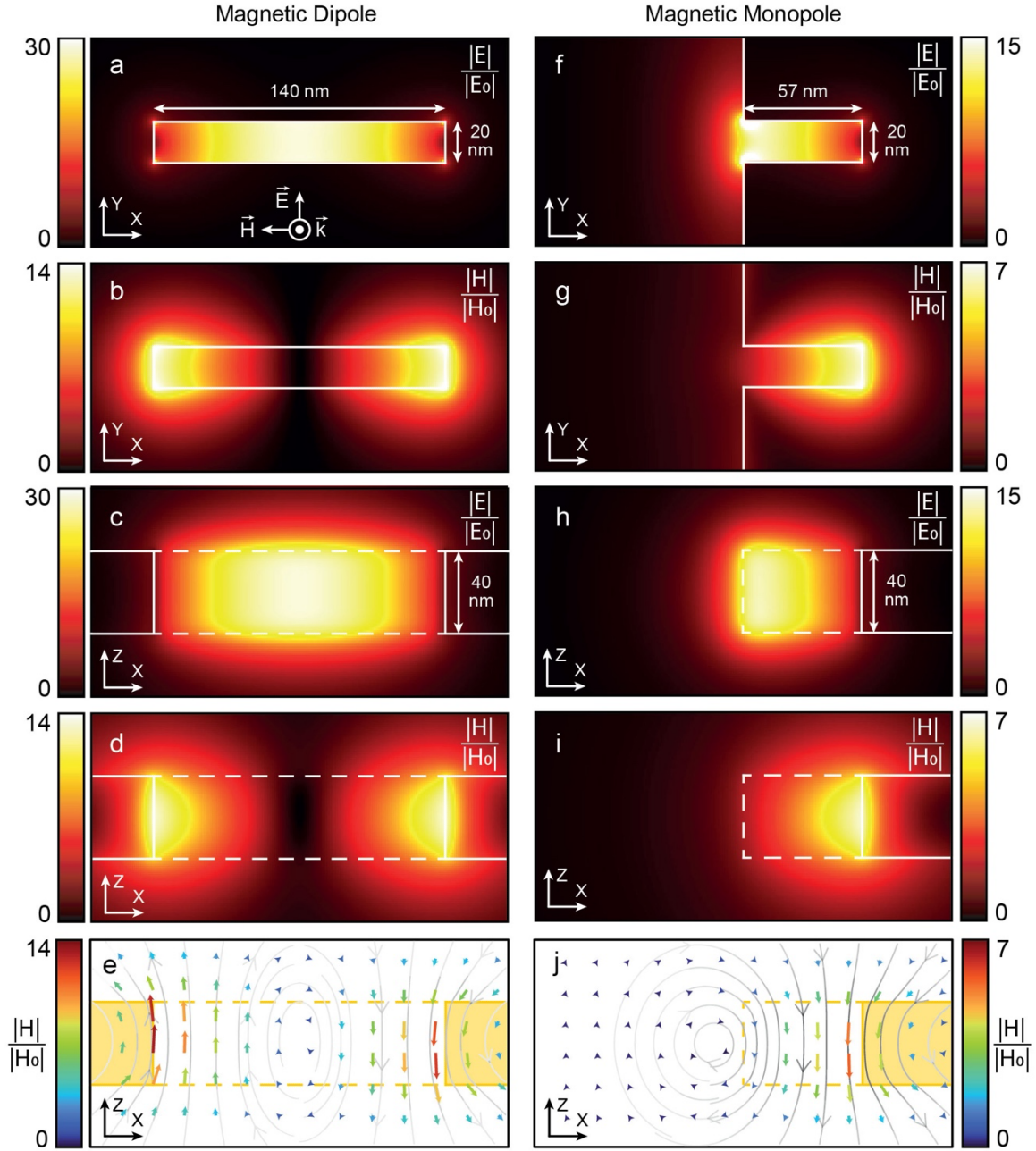

*Figure 4. Spatial distributions of electric and magnetic fields in the near field. Spatial distributions of the (a,c,f,h) electric and (b,d,g,i) magnetic fields for the (a-d) magnetic dipole and (f-i) magnetic monopole antenna in an (a,b,f,g) XY plane at the Z-center of the nanostructures and in a (c,d,h,i) XZ plane at the Y-center of the nanoantennas. Field lines, amplitudes, and vector distributions of the optical magnetic field for e) the magnetic dipole and j) the magnetic monopole in an XZ plane at the Y-center of the structures. The yellow parts indicate the position of the nanoantennas.*

Finally, this magnetic monopole being an electromagnetic antenna, the orientation of the magnetic field lines oscillates in time so that the orientation of the magnetic field changes accordingly at the same frequency. This means that the pole of this antenna must alternate between south and north, for instance. Figure 5 shows this phenomenon. In this figure, the lines, the vector orientation, and the amplitude of the magnetic field over time at six equidistant moments of an optical cycle are shown (see the full video online). As one can see, the magnetic field amplitude oscillates over time, passing through maxima and minima. Also, as expected, the two extrema present in each optical cycle are oriented in opposite ways, one being oriented toward the positive Z, which can be called north, and the other towards the negative Z, which is south. Therefore, we have a north-south fluctuation in the orientation of this pole and, thus, an oscillating magnetic monopole nanoantenna.

It is also important to observe that the field line loops that appear in figures 5 b and e are not those of a dipolar behavior. Instead, they correspond to the electromagnetic wave propagating from the magnetic monopole. For example, in figure 5b, the upward field lines on the left are the reminiscence of the upward magnetic field of figure 5a generated by the monopole and propagating. This propagation can also be observed in figure 5c, where the downward-oriented field lines generated earlier in time in figure 5b in the monopole progressively replace the upward-oriented field lines due to the oscillating character of the monopole. This behavior can be seen very clearly in the video posted online.

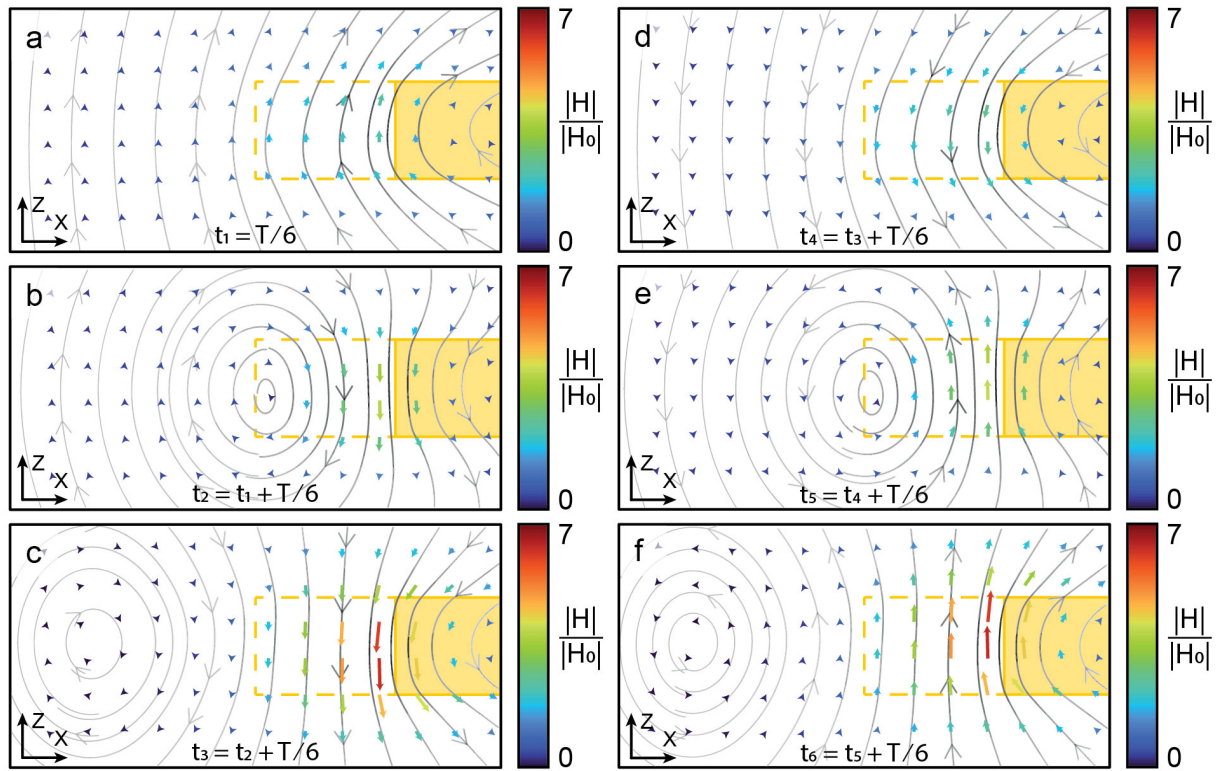

*Figure 5. Temporal study of the **H**-field vectorial distribution. Field lines, amplitudes, and vector distributions of the magnetic field in an XZ plane at the Y-center of the magnetic monopole at different times of a single optical cycle (the time is indicated in each figure). The yellow parts indicate the position of the monopole antenna.*

In conclusion, we have demonstrated that a half-nanoslit carved in a thin semi-infinite gold layer behaves as a magnetic monopole antenna. We have shown that the latter allows generating of a single magnetic hot spot where the optical magnetic field is enhanced. Furthermore, by studying the field lines of this concentration of magnetic energy density, we established that the vectorial distribution corresponded to a single pole, validating the monopole character of this antenna. Finally, by studying the temporal response of this system, we have shown that the magnetic vectorial distribution alternated between north-south orientations over time due to the electromagnetic character of this magnetic monopole, thus allowing this antenna to radiate its energy efficiently in the far field.

In addition to this antenna's new model for magnetic monopoles, it also opens the way to generating intense magnetic hot spots, pure of any electric field, to couple magnetic light to matter[47] efficiently. In particular, topics strongly dependent on the magnetic component of light will directly benefit from this type of system. Finally, although the present study is done at optical wavelengths, the extension of this type of antenna to lower frequencies opens new possibilities in terms of applications, such as magnetic resonance imaging.[48]

**Method section:**

The simulations were performed using the FDTD (Finite Difference Time Domain) software Lumerical. The total computational window was chosen to be 2x2x1.4 µm$^3$ along the X, Y, and Z directions, respectively. A 1 nm fine mesh of 200x100x100 nm$^3$ in the X, Y, and Z directions was used for the area surrounding the dipole and magnetic monopole nanoantennas. An incident plane wave (λ = 800 nm), polarized perpendicular to the long axis of the nanoslits (i.e., along Y) generated at 700 nm below the gold layer, propagates in the positive Z direction.

The electromagnetic fields presented in this manuscript correspond only to the fields generated and scattered by the antenna. For this purpose, the incident light incident on the nanostructures has been subtracted to represent only the signal emitted by the antennas. These incident waves correspond to the electromagnetic fields resulting from the same simulations as the magnetic dipole and monopole (the same infinite and semi-infinite gold layers) but without the presence of the antennas.

**Research funding:** This work is supported by the Agence National de la Recherche (ANR-20-CE09-0031-01), the Institut de Physique du CNRS (Tremplin@INP 2020) and the China Scholarship Council.

**Conflict of interest statement:** The authors declare no conflict of interest regarding this article.